\documentstyle[prl,aps,epsf,multicol]{revtex}
\begin{document}
\title{\bf Exact Maximal Height Distribution of Fluctuating Interfaces }
\author { Satya N. Majumdar $^1$ and Alain Comtet $^2.^3$}
\address{
{\small $^1$Laboratoire de Physique Theorique (UMR C5152 du CNRS), Universit\'e Paul
        Sabatier, 31062 Toulouse Cedex. France}\\
{\small $^2$Laboratoire de Physique Th\'eorique et Mod\`eles Statistiques,
        Universit\'e Paris-Sud. B\^at. 100. 91405 Orsay Cedex. France}\\
{\small $^3$Institut Henri Poincar\'e, 11 rue Pierre et Marie Curie, 75005 Paris, France}}
\date{\today}

\maketitle

\begin{abstract}
We present an exact solution for the distribution $P(h_m,L)$ of the maximal
height $h_m$ (measured with respect to the average spatial height) in
the steady state of a fluctuating Edwards-Wilkinson interface 
in a one dimensional
system of size $L$ with both periodic and free boundary conditions.
For the periodic case, we show that $P(h_m,L)=L^{-1/2}f\left(h_m L^{-1/2}\right)$
for all $L>0$ where the function $f(x)$ is the Airy distribution function 
that describes the probability density of the area under a Brownian excursion
over a unit interval. For the free boundary case, the same scaling holds but
the scaling function is different from that of the periodic case.
Numerical simulations are in excellent agreement with our analytical results.
Our results provide an exactly solvable case for the distribution
of extremum of a set of {\it strongly correlated} random variables.

\noindent

\end{abstract}

\begin{multicols}{2}

Fluctuating interfaces are amongst the most well studied nonequilibrium 
systems due to their simplicity as well as numerous practical applications
in systems such as growing crystals, molecular beam epitaxy,
fluctuating steps on metals and growing bacterial colonies\cite{Review}.  
While the past studies mostly focused on the scaling properties of the surface 
roughness characterized by the average width of the surface height\cite{Review}, 
the more recent theoretical and experimental studies have 
dealt with a variety of other important characteristics of a fluctuating
interface. These include the distribution of the width of heights in the steady state
\cite{Width}, the statistics of first-passage events 
or persistence\cite{Pers,Maryland}, the density of local maxima or minima of heights\cite{Lcm},  
the distribution of the spatially averaged height\cite{Ldf} 
as well as the distribution of height at any fixed point in space\cite{PS} 
in growing one dimensional Kardar-Parisi-Zhang (KPZ) interfaces\cite{KPZ},  
the cycling effects\cite{Cycling}, the distribution of extremal Fourier intensities\cite{GHPZ}
etc.

Recently Raychaudhuri et. al. \cite{RCPS} studied a different characteristic,
namely the global maximal relative height (MRH) (measured with respect to the spatially
averaged growing height) of a fluctuating interface. This is an important observable
for two principal reasons. First, it has important technogical significance
such as in batteries where a short circuit occurs when the highest point
of a metal surface on one electrode reaches the opposite one\cite{RCPS}.
Secondly, the maximal height is an extreme observable measuring a rare event.
While the extreme value statistics is well
understood for a set of {\em independent} random variables\cite{Extreme}, only
recently physicists have been paying attention to the distribution of
the extremum of a set of {\em correlated} random variables, as this question
is appearing increasingly frequently in a number of problems ranging from
disordered systems\cite{DS} to various problems in
computer science such as growing search trees\cite{Trees} and networks\cite{Net}.
In a fluctuating interface, the heights are strongly correlated and hence a
knowledge of the distribution of their maximum (or minimum) would provide
important insights into this important general class of extreme value problems
where the random variables are correlated.

In Ref. \cite{RCPS}, the authors argued quite generally that the MRH $h_m$ of
an interface in its stationary state in a finite system of size $L$ scales
as the roughness of the surface, $h_m \sim L^{\alpha}$ for large $L$, where
$\alpha$ is the roughness exponent. This indicates that the normalized probability
density (pdf) of $h_m$ has a scaling form, $P(h_m,L)\sim L^{-\alpha}f\left( 
h_m/L^{\alpha}\right)$.
This was demonstrated numerically in \cite{RCPS} for a one dimensional lattice
model belonging to the Edwards-Wilkinson (EW) universality class\cite{Edwards},
where $\alpha=1/2$.
Further, it was argued that the scaling function $f(x)$ is sensitive 
to the boundary conditions\cite{RCPS}. 

In this Letter, using simple path integral techniques we present an exact solution of the scaling 
function 
$f(x)$ for the
one dimensional EW model, both for the periodic and the free boundary conditions.
For the periodic boundary case, we show that the scaling function $f(x)$ is the so called
Airy distribution function (not to be confused with the Airy function) which
is the pdf of the area under a Brownian excursion
over a unit interval and has been well studied in the mathematics 
literature\cite{Darling,Louchard,Takacs,PW}. 
We also calculate exactly the corresponding scaling function for the free boundary condition
and show that it is different from the periodic case.
All the moments of $h_m$ are also computed exactly for both the boundary conditions. 
These results are in excellent agreement with the simulation results obtained by
the numerical integration of the discretized $1$-d EW equation.
Our results thus
provide an exactly solvable case for the distribution of the extremum of a set of
{\it strongly correlated} random variables.

Our starting point is the one dimensional EW model\cite{Edwards} which prescribes a linear
evolution equation for the height $H(x,t)$, 
\begin{equation}
{{\partial H(x,t)}\over {\partial t}}= {{\partial^2 H(x,t)}\over {\partial x^2}}+\eta(x,t),
\label{ew1}
\end{equation}
where $\eta(x,t)$ is a Gaussian white noise with zero mean and a correlator, $\langle 
\eta(x,t)\eta(x't')\rangle= 2 \delta(x-x')\delta(t-t')$. The equation (\ref{ew1}) has a soft
(zero wave vector) mode since the spatially averaged height ${\overline {H(x,t)}}= \int_0^{L} 
H(x,t)dx/L$ keeps on growing with time (typically as $\sqrt{t/L}$) even in a finite system
of size $L$. Hence, it is useful to subtract this zero mode from the height and define the relative 
height, $h(x,t)= H(x,t)-{\overline {H(x,t)}}$ whose distribution then reaches a stationary state
in the long time limit in a finite system. Note that, by definition, 
\begin{equation}
\int_0^{L} h(x,t) dx =0.
\label{k0}
\end{equation}
We will see later that this constraint of zero total area under the relative height $h$ plays an 
important role in determining
the MRH distribution. All the other nonzero modes of $h$ evolve identically
as those of the actual height $H$.

We first consider the periodic boundary condition, $h(0)=h(L)$. In this case, one can decompose the 
relative height $h(x,t)$ into a Fourier series, $h(x,t)= \sum_{m=-\infty}^{\infty} {\tilde 
h}(m,t)e^{2\pi i mx/L}$. Substituting this in Eq. (\ref{ew1}), one finds that
different nonzero Fourier modes decouple from each other and one can easily calculate
any correlation function. In particular, it is easy to see that the height $h(x,t)$ at any 
given point converges to a stationary Gaussian distribution as $t\to \infty$,
$P_{\rm st}(h)= e^{-h^2/{2w^2}}/{\sqrt{2\pi w^2}}$ where the width
$w(L)=\sqrt{\langle h^2\rangle}= \sqrt{L/12}$ for all $L$. Moreover,
one can also show that $\langle \partial_x h \partial_{x'} h\rangle \to \delta(x-x')-1/L$
in the stationary state. The local slopes $\partial_x h$ are thus uncorrelated
except for the overall constraint due to the periodic boundary condition, $\int_0^L dx\, 
\partial_x h =0$ that gives rise to the residual $1/L$ term. These informations can 
be collected together
to write the joint probability distribution of the heights (multivariate Gaussian distribution) in 
the stationary state as,
\begin{eqnarray}
P\left[\{h\}\right]&=& C(L) \, e^{-{1\over {2}}\int_0^{L}d\tau (\partial_{\tau} h)^2}\,
\delta\left[h(0)-h(L)\right]\times \nonumber \\
&\times&  \delta\left[ \int_0^{L} h(\tau)d\tau\right],
\label{mpbc}
\end{eqnarray}
where $C(L)$ is a normalization 
constant and the two delta functions take care respectively of the periodic boundary condition
and the zero area constraint in Eq. (\ref{k0}). The constant $C(L)= \sqrt{2\pi} L^{3/2}$ can
be evaluated exactly by integrating Eq. (\ref{mpbc}) over all heights and setting it to 
unity\cite{Details}. One can check that if one integrates out all the heights in Eq. 
(\ref{mpbc}) except at one point, one recovers the single point stationary height distribution 
mentioned before.

We next define the cumulative distribution of the MRH, $F(h_m,L)= {\rm 
Prob}\left[ {\rm max}\{h\}<h_m, L\right]$. The pdf of the MRH is simply the derivative,
$P(h_m,L)= {{\partial F(h_m,L)}\over {\partial h_m}}$.
Clearly $F(h_m,L)$ is also the probability that the heights at all points
in $[0,L]$ are less than $h_m$ and can be formally written using the measure in Eq. (\ref{mpbc})
as a path integral,
\begin{eqnarray}
F(h_m, L)&=& C(L) \int_{-\infty}^{h_m} d u\, \int_{h(0)=u}^{h(L)=u} {\cal D} h(\tau)\, 
 e^{-{1\over {2}}\int_0^{L}d\tau (\partial_{\tau} h)^2} \times \nonumber \\
&\times &\delta\left[ \int_0^{L} h(\tau)d\tau\right] I(h_m,L),
\label{cum1}
\end{eqnarray}
where $I(h_m,L)= \prod_{\tau=0}^{L}\theta(h_m-h(\tau))$ is an indicator function 
which is $1$ if all the heights are less than $h_m$ and zero otherwise.
All the paths inside the path integral propagate from its initial value $h(0)=u$
to its final value $h(L)=u$, where $u\le h_m$ (since by definition $h_m$ is the maximum).
A change of variable, $y(\tau)= h_m-h(\tau)$ and $v=h_m-u$ in the path
integral in Eq. (\ref{cum1}) gives,
\begin{eqnarray}
F(h_m, L)&=& C(L) \int_{0}^{\infty} d v\, \int_{y(0)=v}^{y(L)=v} {\cal D} y(\tau)\,
 e^{-{1\over {2}}\int_0^{L}d\tau (\partial_{\tau} y)^2} \times \nonumber \\
&\times &\delta\left[ \int_0^{L} y(\tau)d\tau -A \right] I(h_m,L),
\label{cum2}
\end{eqnarray}  
where $I(h_m,L)= \prod_{\tau=0}^{L} \theta(y(\tau))$ and $A=h_m L$. Note that
$h_m$ appears only through $A$ in the delta function, and hence
$F(h_m, L) = {\cal F}(A, L)$. In subsequent
calculations, we will keep a general $A$ in Eq. (\ref{cum2}) and will finally
use $A=h_m L$. Next we take the Laplace transform with respect to $A$ in Eq. (\ref{cum2})
and identify the quantity inside the exponential as the action 
corresponding to a single particle quantum Hamiltonian, ${\hat H}\equiv -{1\over {2}} { 
{\partial^2}\over {\partial y^2}} + V(y)$,
where $V(y) = \lambda y $ for $y>0$
and $V(y)=\infty$ for $y\le 0$. The latter condition takes care of the indicator function.
Using the standard bra-ket notation we get,
\begin{eqnarray}
\int_0^{\infty} {\cal F}(A,L)e^{-\lambda A}dA &=& C(L) \int_{0}^{\infty} dv <v|e^{-{\hat H}L}|v> 
\nonumber \\
&=& C(L)\, {\rm Tr}\left[ e^{-{\hat H}L} \right],
\label{trace1}
\end{eqnarray}      
where ${\rm Tr}$ is the trace. 
Thus our problem is reduced to calculating just the eigenvalues of the above
Hamiltonian $\hat H$ which has only bound states and hence discrete eigenvalues.
Solving the Schr\"odinger equation, one finds that the wavefunction (up to a
normalization constant) is simply $\psi_{E}(y)= Ai\left[(2\lambda)^{1/3}(y-E/\lambda)\right]$
where $Ai(z)$ is the standard Airy function\cite{AS}. This wavefunction must vanish at $y=0$
which determines the discrete eigenvalues,  
$E_k = \alpha_k \lambda^{2/3} 2^{-1/3}$ for $k=1,2\dots$, where $\alpha_k$'s
are the magnitude of the zeros of $Ai(z)$ on the negative real axis.
For example, one has $\alpha_1=2.3381\dots$, $\alpha_2=4.0879\dots$, $\alpha_3=5.5205 \dots$ etc. 
Upon formally inverting the Laplace transform in Eq. (\ref{trace1}) 
and putting $A=h_m L$ we find
\begin{equation}
F(h_m,L)= \sqrt{2\pi} L^{3/2} \int_{-i\infty}^{+i \infty} {{d\lambda}\over {2\pi i}} e^{\lambda 
h_m L} \sum_{k=1}^{\infty} e^{-\alpha_k \lambda^{2/3} 2^{-1/3} L}.
\label{ilt1}
\end{equation}
Taking derivative with respect to $h_m$ in Eq. (\ref{ilt1}) and making a change
of variable, $\lambda= sL^{-3/2}$, we arrive
at our main result,
$P(h_m,L)= L^{-1/2} f\left(h_m L^{-1/2}\right)$ for all $L$, where the Laplace transform of
$f(x)$ is given by
\begin{equation}
\int_0^{\infty} f(x) e^{-sx} dx = s\sqrt{2\pi} \sum_{k=1}^{\infty} e^{-\alpha_k s^{2/3}2^{-1/3}}.
\label{lt1}
\end{equation}

Interestingly, the right hand side of Eq. (\ref{lt1}) turns out precisely to be the Laplace transform 
of the pdf of the area under a
Brownian excursion over a unit interval\cite{Takacs}.
A Brownian excursion over the interval
$[0,1]$ is simply a Brownian motion pinned at zero at the two ends of
the interval and  conditioned 
to stay positive in between. Inverting the Laplace transform in Eq. (\ref{lt1}) one
obtains $f(x)$, known as the Airy distribution function\cite{Takacs},
\begin{equation}
f(x)= {{2\sqrt{6}}\over {x^{10/3}}}\sum_{k=1}^{\infty} e^{-b_k/x^2} b_k^{2/3}
U(-5/6, 4/3, b_k/x^2),
\label{fx1}
\end{equation}
where $b_k= 2\alpha_k^3/{27}$ and $U(a,b,z)$ is the confluent hypergeometric function\cite{AS}.
In Fig. 1, we have plotted $f(x)$ in Eq. (\ref{fx1}) using the Mathematica
and compared it with the numerical scaling function generated by collapsing
the data for $3$ different system sizes obtained by numerically integrating  
the space-time discretized form of Eq. (\ref{ew1}). 
Evidently the agreement is very good. 

It is easy to obtain the small $x$ behavior of $x$ from Eq. 
(\ref{fx1}), since only the $k=1$ term dominates as $x\to 0$. Using $U(a,b, z)\sim z^{-a}$ for 
large $z$, we get as $x\to 0$,
\begin{equation}
f(x)\to {8 \over {81}} \alpha_1^{9/2} x^{-5}\,\exp\left[-{{2\alpha_1^3}\over {27 x^2}}\right].
\label{es1}
\end{equation}
This essential singular tail near $x\to 0$ was conjectured in \cite{RCPS} based
on numerics, though the exact form was missing. The asymptotic behavior at large $x$ is more 
tricky to derive\cite{Yor} from
Eq. (\ref{fx1}). Even the calculation of moments from Eq. (\ref{lt1}) is rather nontrivial. 
However, it is possible
to write down an exact recursion relation for the moments\cite{Louchard,Takacs} and using 
these results, we get $\langle h_m^{n}\rangle = M_n L^{n/2}$ where $M_0=1$, $M_1=\sqrt{\pi/8}$,
$M_2=5/12$, $M_3= 15\sqrt{\pi}/{64\sqrt{2}}$, $M_4= 221/1008$ etc. Only the 
second moment $\langle 
h_m^{n}\rangle=5L/12$ was computed before in \cite{RCPS} by using a different method.
Finally, one finds that for large $n$, $M_n \sim (n/12e)^{n/2}$. Substituting
an anticipated large $x$ decay of the form, $f(x)\sim \exp[- a x^b]$ in
$M_n=\int_0^{\infty} f(x) x^{n}dx$, we get $M_n\sim (n/{abe})^{n/b}$ for
large $n$. Comparing this with the exact large $n$ behavior of $M_n$ we get $a=6$ and $b=2$, 
indicating $f(x)\sim \exp[-6x^2]$ as $x\to \infty$.

There is an alternative elegant probabilistic derivation of the above result 
which we outline briefly. It proceeds by establishing
the equivalence,
\begin{equation}
h(x)\equiv B(x)-{1\over {L}}\int_0^{L} B(\tau)d\tau,
\label{pb1}
\end{equation}
where $h(x)$ is the stationary EW interface with periodic boundary condition, $B(x)$
is a Brownian bridge (a Brownian motion such that $B(0)=B(L)=0$) and $\equiv$
means that the left hand side (lhs) has the same probability distribution as the right hand side 
(rhs).  First, by construction the rhs satisfies the area constraint in Eq. (\ref{k0}).
Secondly, both the lhs and rhs of Eq. (\ref{pb1}) are Gaussian variables
and hence to establish the equivalence in Eq. (\ref{pb1}), it is sufficient to show
that their respective two-point correlators are identical. For example,
one finds\cite{Details} from Eq. (\ref{ew1}) that in the stationary state, $\langle h(x)h(x')\rangle
=[L^2/6-L|x-x'| + (x-x')^2]/{2L}$ for all $L$. Similarly, one can calculate
the two-point correlator of the rhs using the representation,
$B(\tau)= x(\tau)- \tau x(L)/L$, where $x(\tau)$ is ordinary Brownian motion
starting at $x(0)=0$ and with a correlator, $\langle x(\tau)x(\tau')\rangle = {\rm min}(\tau,
\tau')$. This representation guarantees that $B(0)=B(L)=0$.
We find that the two-point correlator of the rhs is exactly the same as
$\langle h(x)h(x')\rangle$. This establishes
the equivalence in law in Eq. (\ref{pb1}) rigorously. Hence, the maximum of $h(x)$
will have the same distribution as the maximum of the rhs of Eq. (\ref{pb1}) which,
incidentally, was computed by Darling in the context of statistical data analysis
and he found\cite{Darling} exactly the same Laplace transform
as in Eq. (\ref{lt1}).

We next consider the free boundary condition where the two ends of the interface
are held free. In this case, the joint distribution
of heights in the stationary state is given by the same formula as in Eq. (\ref{mpbc}),
except without the delta function $\delta\left[h(0)-h(L)\right]$.  This changes
the normalization constant to $C(L)=L$. However, unlike the simple trace
in the periodic case in Eq. (\ref{trace1}), the Laplace transform in the free
case turns out to be more complicated\cite{Details}. Omitting details,  
we find the same scaling as in the periodic case, $P(h_m, L)= L^{-1/2} f\left(h_m 
L^{-1/2}\right)$, though the scaling function has a different Laplace transform
${\tilde f}(s)=\int_0^{\infty} f(x) s^{-sx} dx$,
\begin{equation}
{\tilde f}(s) = s^{2/3} 2^{-1/3}\sum_{k=1}^{\infty} C(\alpha_k) e^{-\alpha_k 
s^{2/3}2^{-1/3}},
\label{lt2}
\end{equation} 
where $C(\alpha)= [\int_{-\alpha}^{\infty} Ai(z)dz]^2/[Ai'(-\alpha)]^2$ and
$Ai'(z)=dA_i(z)/dz$. This Laplace transform
does not seem to have appeared before in the mathematics literature. One can again
express the function $f(x)$ in terms of the confluent hypergeometric function\cite{Details},
a Mathematica plot of which is shown in Fig. 1 that matches well with the numerical 
simulations.
The small $x$ behavior can again be found easily and we get,
\begin{equation}
f(x)\to {{2\sqrt {2}}\over {27\sqrt{\pi}}} C(\alpha_1) \alpha_1^{7/2} x^{-4} 
\exp\left[-{{2\alpha_1^3}\over {27 
x^2}}\right],
\label{es2}
\end{equation}
where $C(\alpha_1)=3.30278\dots$, evaluated using the Mathematica. Thus
the function $f(x)$ decays slightly faster as $x\to 0$ compared
to the periodic case in Eq. (\ref{es1}).
We also calculated the moments exactly\cite{Details}, $\langle h_m^n\rangle
=\mu_n L^{n/2}$ where $\mu_0=1$, $\mu_1=\sqrt{2/\pi}$, $\mu_2=17/24$, $\mu_3= 
123\sqrt{2}/{140\sqrt{\pi}}$
etc. We found that for large $n$, $\mu_n\sim [n/3e]^{n/2}$ which provides the 
asymptotic large $x$ tail of $f(x)$, $f(x)\sim \exp[-3x^2/2]$ that 
falls off less rapidly than the periodic case where $f(x)\sim \exp[-6x^2]$.
\begin{figure}
\narrowtext\centerline{\epsfxsize\columnwidth \epsfbox{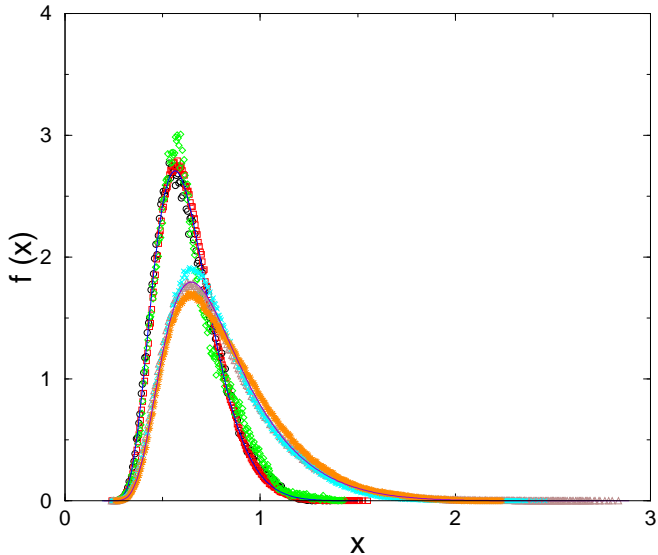}}
\caption{The scaling function $f(x)$ for the MRH distribution for both
the periodic (the top $4$ curves) and the free (the lower $4$ curves)
boundary conditions. In both cases, the numerical curves (shown by symbols) are obtained
by collapsing the data from the numerical integration of the 
space-time discretized form of the EW equation (\ref{ew1})
for $3$ system sizes $L=256$, $L=384$ and $L=512$. They are compared to
the corresponding analytical scaling functions (solid lines).} 
\end{figure}

\vspace{-0.2cm}

To conclude, we note that apart from the theoretical interests as a solvable model, many 
experimental systems are well described by the $1$-d EW equation
(\ref{ew1}). Examples include, amongst others,  the high-temperature step fluctuations in 
Si(111)-Al 
surfaces\cite{Maryland,Giesen} and the displacements of nonmagnetic particles in dipolar chains
at low magnetic field\cite{THF}. Besides, the displacements 
of beads in a polymer chain with harmonic interaction (the Rouse model\cite{Rouse})
also evolve via the $1$-d EW equation. Thus our results are relevant
in these systems and it would be interesting to see if the MRH 
distribution can be measured experimentally in such systems.

We thank Y. Shapir, S. Raychaudhuri and C. Dasgupta for useful correspondence and discussions. 

\end{multicols}

\end{document}